\documentclass[12pt,preprint]{aastex} 
\usepackage{psfig}

\shorttitle{The Nature of the SS 433 Binary}
\shortauthors{Lopez et al.}

\begin{document}
\title{Determining the Nature of the SS 433 Binary from an X-ray Spectrum During Eclipse}
\author{Laura A. Lopez\altaffilmark{1,2,3}, Herman L. Marshall\altaffilmark{4}, Claude R. Canizares\altaffilmark{4}, Norbert S. Schulz\altaffilmark{4}, Julie F. Kane\altaffilmark{5}}
\altaffiltext{1}{Department of Astronomy and Astrophysics, University of California Santa Cruz, 355 Interdisciplinary Sciences Building, 1156 High Street, Santa Cruz, CA 95064, USA; lopez@astro.ucsc.edu.}
\altaffiltext{2}{Institute for Advanced Study, School of Natural Sciences, Einstein Lane, Princeton, NJ 08540}
\altaffiltext{3}{NSF Graduate Research Fellow}
\altaffiltext{4}{Department of Physics and Kavli Institute for Astrophysics and Space Research, Massachusetts Institute of Technology, Cambridge, MA 02139; hermanm@space.mit.edu, crc@space.mit.edu, nss@space.mit.edu}
\altaffiltext{5}{Department of Earth, Atmospheric, \& Planetary Science, Massachusetts Institute of Technology, Cambridge, MA 02139; jfkane@mit.edu}

\begin{abstract}
We test the physical model of the relativistic jets in the galactic X-ray binary SS 433 proposed in our previous paper using additional observations from the {\it Chandra} High Energy Transmission Grating Spectrometer. These observations sample two new orbital/precessional phase combinations. In the observation near orbital phase zero, the H- and He-like Fe lines from both receding and approaching jets are comparably strong and unocculted while the He-like Si line of the receding jet is significantly weaker than that of the approaching jet. This condition may imply the cooler parts of the receding jet are eclipsed by the companion. The X-ray spectrum from this observation has broader emission lines than obtained in Paper I that may arise from the divergence of a conical outflow or from Doppler shift variations during the observation. Using recent optical results, along with the length of the unobscured portion of the receding jet assuming adiabatic cooling, we calculate the radius of the companion to be \hbox{$9.6\pm1.0$ $R_{\sun}$}, about one third of the Roche lobe radius. For a main sequence star, this corresponds to a companion mass of $35\pm7$ $M_{\sun}$, giving a primary source mass of $20\pm5$ $M_{\sun}$. If our model is correct, this calculation indicates the compact object is a black hole, and accretion occurs through a wind process. In a subsequent paper, we will examine the validity of the adiabatic cooling model of the jets and test the mode of line broadening. 
\end{abstract}

\keywords{X-ray sources, individual: SS~433}

\section{Introduction}

The Galactic X-ray binary SS 433 is the only known astrophysical jet system in which relativistically red- and blue-shifted lines are observed from high-Z elements. Lines from the H-Balmer series were detected (Margon et al. 1977) and modeled kinematically (Abell \& Margon 1979; Fabian \& Rees 1979; Milgrom 1979) as emission from opposing jets emerging from the vicinity of a compact object that precesses about an axis inclined to the line of sight. The jet velocity $v_j$ is 0.260$c$, and its orientation precesses with a 162.5 day period in a cone with half-angle 19.85\arcdeg\ about an axis which is 78.83\arcdeg\ to the line of sight (Margon \& Anderson 1989). The lines from the jet are Doppler shifted with this period so that the maximum
redshift is about 0.15 and the maximum blueshift is about -0.08. The X-ray source undergoes eclipses at a 13.0820 day period. Assuming uniform outflow, the opening angle of the jet is 5\arcdeg\ based on the widths of the optical lines (Begelman et al. 1980). For more details of the optical spectroscopy, see the review by
\citet{margon84}.
 
SS 433 has radio jets oriented at a position angle of 100\arcdeg\ (east of north) which show an oscillatory pattern that arises from helical motion of material flowing along ballistic trajectories \citep{hjellming}.
Using VLBI observations of ejected knots, \citet{vermeulen} independently confirmed the velocity of the jet and
the kinematic model and also derived an accurate distance: 4.85 $\pm$ 0.2 kpc. We adopt this distance throughout our analysis. The radio jets extend from the milliarcsec scale to several arcseconds, a physical range of $10^{15-17}$ cm from the core.  The optical emission lines, however, originate from a smaller region, $< 3 \times 10^{15}$ cm across, based on light travel time arguments \citep{dm80}.

Using {\em HEAO-1}, \cite{marshall79} were the first to demonstrate that SS 433 is an X-ray source. The {\em HEAO-1} continuum was sufficiently modeled as thermal bremsstrahlung with \hbox{$kT = 14.3$ keV}, and emission due to Fe-K was detected near 7 keV. Based on {\em Ginga} observations, \citet{brinkmann91} concluded that $kT$ is $\approx$30 keV, while \citet{kotani96} obtained a value of 20 keV based on the ratio of  Fe {\sc xxv} and Fe {\sc xxvi} line fluxes from {\em ASCA}. The X-ray emission may originate from the base of the jets which adiabatically expand and cool until $kT$ drops to $\approx$100 eV and become thermally unstable (Brinkmann et al. 1991; Kotani et al. 1996). \cite{kotani96} found that the redshifted Fe {\sc xxv} line was fainter relative to the blueshifted line than expected from Doppler intensity conservation. Consequently, they concluded that the redward jet must be obscured by neutral material in an accretion disk.  Although the {\em ASCA} spectra showed emission lines that were previously unobserved, the lines were not resolved spectrally and lines below 2 keV were difficult to identify.

\citet{marshall02} [hereafter, Paper I] observed SS 433 with the High Energy Transmission Grating Spectrometer (HETGS; Canizares et al. 2005) of the {\it Chandra} X-ray Observatory. This observation resolved the X-ray lines, detected fainter lines than previously observed, and measured lower energy lines than could be readily detected in the {\em ASCA} observations. Additionally, the HETGS spectrum of SS 433 showed many emission lines previously identified in the {\em ASCA} observations \citep{kotani94}. In Paper I, the X-ray spectrum was dominated by thermal emission from the jets. Additionally, using the Si {\sc xiii} triplet, the electron density of the jet was determined to be 10$^{14}$ cm$^{3}$, facilitating an estimate for the size of the jet, (2--20)$\times$10$^{10}$ cm.  

In Paper I, blue-shifted lines dominated the spectrum and the red-shifted lines were all relatively weak by comparison, so most work focused on modeling the blue jet. The jet velocity was determined very accurately due to the small scatter of individual line Doppler shifts and the narrowness of their profiles, indicating that all line-emitting gas flows at the same speed. The jet bulk velocity was $\beta$$c$, where $\beta$=$0.2699\pm0.0007$. This jet velocity is larger than the model velocity inferred from optical emission lines by $2920\pm440$ km s$^{-1}$. Gaussian fits to the emission lines were all consistent with the same Doppler broadening: \hbox{$1700\pm80$ km s$^{-1}$} (FWHM). Relating this broadening to the maximum velocity due to beam divergence, the opening angle of the jet was determined to be $1.23^{\circ}\pm0.06^{\circ}$. 

Here, we present new observations of SS 433 using the HETGS to test the physical model of the jets proposed in Paper I. The new observations were taken at two different combinations of precessional and orbital phases, offering contrasting views of the jets. In particular, one observation was taken during eclipse. The goal of the observation was to determine which parts of the jets would be occulted due to this eclipse. The results of the optical observations presented by \cite{gies1}, along with the length of the jets determined assuming an adiabatic cooling model, are utilized to measure the size and mass of the companion. This analysis will set a restriction on the mass of the compact object and ultimately support the hypothesis that the primary source is a black hole. Our conclusions rely on the postulate that adiabatic cooling dominates over radiative losses, and we will discuss the limitations of this model in $\S$5.1. 

Previous X-ray observations of the SS 433 system during eclipse \citep{brinkmann91,stewart} also displayed weak emission lines from the receding jet. This result has led some to believe the entire inner region of the jets was eclipsed. However, our spectrum during occultation shows numerous emission lines from the receding jet. Particularly, the presence of strong, highly-ionized Fe lines indicates only the low-energy portion of this jet was eclipsed.
\section{Observations and Data Reduction}

SS 433 was observed with the HETGS on 2000 November 21 and 2001 March 16 for 24 ks each, observation IDs 1020 and 1019.  The latter observation was taken during eclipse when the companion blocked part of the X-ray continuum.  Based on the precession ephemeris by \cite{ma89} as updated by \cite{gies2}, the angle of the jet to the line of sight during this observation was predicted to be $\alpha$=80.0\arcdeg. From the precession and binary ephemeris by \cite{gladyshev}, the binary phase range was 0.96 to 0.98.  For observation 1020, $\alpha$ was 96.0\arcdeg\ (meaning that the normally appoaching western jet was temporarily pointed farther from the line of sight than than the normally receding eastern jet) and the binary phase range was 0.68-0.70.
\subsection{Imaging}

The zeroth-order image from the three observations (the Paper I observation and the two presented in this paper) is shown in Figure 1 and is distinctly elongated in the east-west direction. The extended X-ray emission was first reported in Paper I. This result is consistent with \cite{fender} who found evidence for outflowing hot plasma in the extended X-ray emission. The scale of emission is 10$^{4}$ larger than the region producing the bulk of the X-ray spectrum, indicating plasma is reheating very far from the compact object. The excess emission in the east-west direction is associated with the arcsecond-scale radio jet, as observed by \cite{hjellming} using the VLA.

Based on the dispersed spectra (because the point source is somewhat piled up, as in Paper I), the X-ray flux from SS 433 was constant during observation 1019 but increased by about 15\% on a time scale of 10000 s in November 2000.
  
\subsection{Spectra}
\label{sec:spectra}

As in Paper I, the spectral data were reduced starting from level 1 data provided by the {\em Chandra} X-ray Center (CXC) using IDL custom processing scripts. The spectra from the November 2000 and March 2001 observations are shown in Figs.~\ref{fig:spectrum} and \ref{fig:spectrum_obs2}. The jet emission lines are much weaker in the November 2000 observation than in March 2001 or Paper I. Consequently, the November 2000 will not be analyzed further in this paper, and we will only discuss results from March 2001 as compared to Paper I. 

Line fluxes for the March 2001 observation are given in Table 1. A similar methodology from the first paper was utilized: lines were fit to the MEG and HEG data jointly in five wavelength ranges after subtracting a polynomial fit to the continuum. Gaussian widths are given in Table 2. Only statistical errors are quoted; systematic uncertainties other than possible line misidentifications are expected to be smaller than 10\%. The lines are twice as broad as those found in the first observation.  
The spectra are shown in Figs.~\ref{fig:spectrum} to ~\ref{fig:spectrum3}. Figure~\ref{fig:spectrum} shows the flux corrected spectrum, combining
the MEG and HEG data with statistical weighting. Similar to the first observation, many broad emission lines and a significant continuum are prevalent. Unlike Paper I where some red jet lines were undetectably weak, every blue jet line has a corresponding red jet line in the March 2001 observation.

For each line, the ID was determined based on the expected shifts of the blue and red jets from the kinematic model and the measured redshift. For the precession phase of our observation, the predicted blue- and red-shifts were 0.011 and 0.063, respectively, using the equation and parameters in Table 1 of \cite{ma89}. Wavelengths of line blends were obtained using wavelengths from the Astrophysical Plasma Emission Database (APED\footnote{See {\tt http://cxc.harvard.edu/atomdb/sources\_aped.html} 
for more information about APED.}) weighted by the relative fluxes of each component based on the models from $\S$5. The S {\sc xv} and Si {\sc xiii} triplets were analyzed as in Paper I. Precise redshifts for these profiles were determined by fitting these blends with several variable width Gaussian components and with fixed rest wavelength values given by APED, as implemented in Interactive Spectral Interpretation System  ({\tt ISIS}\footnote{See {\tt http://space.mit.edu/CXC/ISIS} for more information about ISIS.}). Unlike in Paper I, the red and blue jet lines were prominent enough to facilitate accurate measurement of the redshift for both jets. Two lines were not assigned to either the red or blue jets: a neutral Fe-K line and a Si {\sc i}-K line, both at rest in the observed frame.  
\section{Line Widths and Positions}
\label{sec:linewidths}

Table~\ref{tab:linefluxes} shows that the Doppler shifts of the lines in the blue jet system are consistent with a single velocity to within the uncertainties, as are those in the red jet system. The few deviations are mostly due to blended lines (e.g., line triplets) in each jet and are not likely indicative of intrinsic variations. Furthermore, all deviations are substantially smaller than the observed line widths, which are on the order of $\delta$z = 0.006. The Doppler shift of the blue jet was determined to be $z_b=0.0111 \pm 0.0001$ during this observation (as opposed to the $z_b = -0.0779 \pm 0.0001$ in Paper I). For the red jet, the unblended lines give $z_r = 0.0610 \pm 0.0001$ (compared to the $z_r = 0.1550 \pm 0.0004$ reported in Paper I). 

As in Paper I, we assume the system is comprised of two perfectly opposed jets at an angle $\alpha$ to the line of sight. Then, the Doppler shifts of the blue and red jets are given by 

\begin{equation}
\label{eq:redshift}
 z = \gamma (1 \pm \beta \mu ) - 1
\end{equation}

\noindent
where v$_j = \beta c$ is the velocity of the jet flow,
$\gamma = (1 - \beta^2)^{-1/2}$, and $\mu = \cos \alpha$. As in the first paper, high accuracy redshifts were used to obtain an estimate of the $\gamma$ and $\beta$ by adding the Doppler shifts to cancel the $\beta \mu$ terms:

\begin{equation}
\label{eq:gamma}
\gamma = \frac{z_b + z_r}{2} + 1
\end{equation}

\noindent
giving $\beta = 0.2666 \pm 0.0006$. This value is slightly smaller than the $\beta$ obtained in the first paper, $0.2699 \pm 0.0007$, and is closer to, but still larger than, the jet velocity determined by \cite{ma89} based on the H-$\alpha$ lines. Substituting the value for $\beta$ back into Eq.~\ref{eq:redshift} and solving for $\alpha$ gives the angle of the jet to the line of sight during this observation: $\alpha$ = 84.6\arcdeg $\pm$ 0.1\arcdeg.

All jet lines are clearly resolved. The line widths in Table~\ref{tab:linewidths} are consistent with the weighted average value ($\sigma$) of $1430 \pm 167$ km s$^{-1}$. We note that $\sigma$ is $\approx$2 times the value in Paper I: the observed lines are twice as broad as those reported in Paper I. Marginal evidence exists indicating a trend that the lower energy lines are slightly narrower than average. As found in Paper I, the widths of the red jet lines are consistent with those of the blue jet. 

The line widths are too large to result from thermal broadening for $kT < 10$ keV (see $\S$4) -- 100-200 km s$^{-1}$. The Doppler broadened widths may result from the divergence of a conical outflow \citep{begelman} or from Doppler shift variations during the observation. We will investigate the relationship between broadening and Doppler shift variations in a subsequent paper on a recent HETGS observation of SS 433 (Marshall et al. 2006). The origin of the line broadening does not affect the line flux measurements and the plasma modeling of the next section. 
\section{The Jet Emission Line Fluxes}
\label{sec:models}

To model the X-ray spectra, we use a plasma diagnostic approach where emission lines correspond to specific jet temperatures. In both the red and blue jets, we observe comparable line strengths in Fe, Ca, and Ar. Longward of 4\AA\, the red jet lines become weaker than the blue jet lines. We assume that this effect arises from occultation of the low-energy part of the red jet by the companion star. Although Si and S are present in the red jet, these lines are not as strong as expected given the blue jet line fluxes. Therefore, we model the red jet with comparatively lower emission measures at longer wavelengths. 

As in Paper I, the spectra were modeled using {\tt ISIS} and the APED atomic database of line emissivities and ionization balance. The emission measure distribution was estimated by fitting the line flux data to a multitemperature emission model assuming approximately solar abundances. Starting from the model used to analyze the first observation, a moderately good fit was obtained to the line fluxes with a four-component model for both jets, as given in Table 3. We assumed that the ISM absorption was the same as determined in \hbox{Paper I}: $N_H=(2.07\pm0.07)\times10^{22}$ cm$^{-2}$. See Figs.~\ref{fig:spectrum1}--\ref{fig:spectrum3} for a detailed comparison of the models with the HETGS data. No formal uncertainties are given for the values in Table 3, but by comparison to the analysis (Table 4 of Paper I), we estimate the uncertainties to be $\pm$30\%. 

The abundances of Fe, S, Si, Mg, and Ne in the models for both jets were increased by 70\% relative to H and He in order to reduce the model continuum to match the observed continuum.  In Paper I, the abundances were increased 30\% over the solar values. The continuum fits well over much of the spectrum, but in the 8--12 \AA\ region, the model predicts a continuum that is systematically low by up to a factor of 2. Adding a fifth, low temperature component, however, predicted emission lines that are not observed. 

For the red jet, the same four-temperature model was used as the blue jet. The two low-temperature components were set to zero to model the low-energy portion of the red jet as blocked by the companion due to eclipse. The resulting two-temperature model is given in Table 3. This method produced reasonable agreement with the data except around the red jet Si {\sc xiii} line.  

Given the broader emission lines in the March 2001 observation, the lack of Doppler boosting, and the slightly shorter exposure time compared to the Paper I observation, the analysis of the Si {\sc xiii} triplet did not produce stringent limits on the jet electron density. In the blue jet, the value of the ratio of the forbidden and intercombination lines, $R \equiv f/i = 1.79 \pm 0.70$, is consistent with the value determined in Paper I. However, it is also consistent with the asymptotic value as $n_e$ drops below $10^{12}$ cm$^{-3}$ for a plasma with $T = 5 \times 10^7$ K (where Si {\sc xiii} is formed). We note that the blue jet Si {\sc xiii} triplet was better resolved in the Paper I observation, and the Paper I measured $R$ value corresponded to  $n_e$ $\approx$ $10^{14}$ cm$^{-3}$. This electron density is consistent with the assumption that flow time is much greater than the recombination timescale. The Si {\sc i} fluorescence line is confused somewhat with the Si {\sc xiii} of the red jet, making it difficult to compare its $R$ or $G$ values with those of the blue jet. 

The spectrum is more complex than the model suggests: additional lines may be present at 8.1 \AA\ and 10.1 \AA\ that are not included in the model, and the line at 5.1 \AA\ (S {\sc xv} in the blue jet) appears to be slightly stronger in the model than in the data. The Si {\sc xiii} red jet line and the Fe {\sc xxiii}/{\sc xxiv} blue jet line are much stronger than the model's prediction. We note that Fe {\sc xxiii}/{\sc xxiv} is not detected from the red jet due to eclipse, and the blue jet Fe {\sc xxiii}/{\sc xxiv} discrepancy may arise from deviation from ionization equilibrium at low temperatures in both jets. Almost all the other line fluxes are matched to within 50\%, and the strongest lines are modeled to within 10\%-20\%. We consider the overall fit to be reasonably satisfactory. 

It is clear from Fig.~\ref{fig:spectrum1} that the highly ionized Fe lines are not occulted during the eclipse, whereas lower ionization states of Si are not as strong in the receding jet (Fig.~\ref{fig:spectrum2}) as in the approaching jet. Thus, we interpret this result as indicating the cooler part of the red jet is blocked by the companion, whereas the hotter portions (where the Fe lines are produced) are unocculted. We assume the jet flow is conical and has a constant opening angle and jet velocity (as in Fig. 9 of Paper I) because ions in different temperature ranges have similar line widths and Doppler shifts. Consequently, the electron density $n_{e}$ is proportion to $r^{-2}$, where $r$ is the distance from the apex of the cone. 

Approximate values for the plasma parameters were determined iteratively in about fifty runs of the {\tt ISIS} four-zone model. Once we found a satisfactory model for the blue jet (see Table 3), the next step was to distinguish the physical differences in the red jet. To calculate the range of acceptable emission measures for the red jet in the four-zone model, we varied the emission measures in Zone 2 and Zone 3, while the Zone 1 emission measure was set to zero and the Zone 4 emission measure was set equal to the value of the blue jet. Initially, the second- and third-component emission measures were set to zero, as displayed in Fig.~\ref{fig:range} {\it bottom}.  Then, the goodness of fits were tested for increasing values of the emission measures in zone 2 and zone 3 (for example, Fig.~\ref{fig:range} {\it top}). Fig.~\ref{fig:confidence} displays the fits used to determine approximate confidence limits on $T_x$. 

We define $\zeta$ as the ratio of the emission measures of the unocculted zones of the red and blue jets. If $r_j$ is the distance from the apex of the cone to the center of zone $j$, then $\zeta$ is related to $x_j$, the distance from the origin of the red jet to the point of truncation (the truncation radius) in zone $j$, by

\begin{equation}
\label{eq:EM}
\zeta_j = \frac{EM_{red,j}}{EM_{blue,j}} = \frac{\int_{0.5r_j}^{x_j}n_{e}^2dV}{\int_{0.5r_j}^{1.5r_j}n_{e}^2dV} = \frac{3}{2} (1-\frac{r_j}{2x_j})
\end{equation}

\noindent
using $n_{e} \propto r^{-2}$ and the definition of emission measure. Here, $r_j = 1.83 \times 10^{11}$ cm for the second zone ($j = 2$), and $r_j = 9.19 \times 10^{10}$ for the third zone ($j = 3$) in the adiabatically-cooled jet model from Paper I (see Table~\ref{tab:model}). Solving for $x_j$, 

\begin{equation}
\label{eq:zeta}
x_j = \frac{3 r_j}{2}(3 - 2\zeta)^{-1}.
\end{equation}

\noindent
By visual inspection, we identify a range of reasonable models for the data. Using this method, we obtain a lower-limit emission measure of $5.4 \times 10^{57}$ cm$^{3}$ in the second zone and an upper limit of $1.16 \times 10^{58}$ cm$^{3}$ in the third zone. The lower and upper limits to the truncation radii using  Eq.~\ref{eq:zeta} are $5.1 \times 10^{10}$ cm and $8.5 \times 10^{10}$ cm, respectively. Thus, $x_j = (6.8 \pm 1.0) \times 10^{10}$ cm (1$\sigma$ uncertainties).

Other conditions (besides eclipse) might cause the dimming observed in the cool portion of the red jet. For example, an electron scattering region could exist that linearly increases in optical depth with distance from the disk. Optical depth $\tau$ is related to jet length through the scattering region $l$ by $\tau = \sigma_T n_e l \geq 1$, where $\sigma_T$ is the Thomson scattering cross-section and $n_e$ is the electron density. At the far side of the red jet, $l = 2r \cos \alpha$, where $r$ is the jet length $2 \times 10^{11}$ cm and $\alpha = 84.6$ \arcdeg\ from $\S$3. Therefore, 

\begin{equation}
\label{eq:density}
n_e \geq \frac{1}{2 \sigma_T r \cos \alpha} \geq 3.6 \times 10^{13} \bigg(\frac{r}{2 \times 10^{11} {\rm cm}}\bigg)^{-1}
\end{equation}

\noindent
This electron density is comparable to that of the jet, so it seems unlikely that the jets propogate through such a thick medium. Similar difficulties arise when considering cold, neutral material. We approximate the path length through the medium to be \hbox{$4 \times 10^{10}$ cm}. For intensity to drop by a factor of 2 for 2 keV photons, a column density of \hbox{$2 \times 10^{22}$ cm$^{-2}$} is necessary along this path length. Thus, the density of neutral material should be of the order \hbox{5$\times 10^{11}$ particles/cm$^{3}$}. This density is implausibly high, and we discount the possibility of either electron scattering by an ionizing gas or absorption by a cold, neutral gas. 

\section{Discussion} 
\label{sec:discussion}

\subsection{Examining the Adiabatic Expansion Model}
\label{subsec:adiabatic}

The thermal evolution of the jet depends on the relative dominance of two cooling terms: adiabatic expansion and radiative losses \citep{kotani96}. In the first HETGS observation, the emission measure distribution $EM(T)$ was predicted adequately from the adiabatic model. If adiabatic cooling dominates, then $T \propto r_{j}^{-4/3}$ for a nonrelativistic gas. Line profile measurements indicate the jet is a conical, uniform velocity flow (see $\S$\ref{sec:models}), and $n_en \propto r^{-4}$. Thus, $EM(T) \propto r_{j}^{-1} \propto T^{3/4}$. The adiabatic model has shortcomings, as shown in Paper I. Examination of Table ~\ref{tab:model} reveals the high-temperature point is too low and the Zone 1 emission measure is too high. The high-temperature line emission could be reduced by Comptonization in the highest-density portion of the jet. The disagreement in the lowest-temperature zone may be attributed to statistical uncertainties of order 30\% (consistent with Paper I) in detecting long wavelength lines. Here, we will explore other possible effects influencing the emission measure. 

The model fits require that the abundances for metals that dominate the line emission (Fe, S, Si, Mg) have changed from 1.3$\times$ solar to 1.7$\times$ solar between observations. If we assume a larger abundance, such as 2.0$\times$ solar and add a fifth spectral component, the continuum increases dramatically overall. In this model, the new component is thermal bremsstrahlung with sufficiently high temperature to strip detectable ions, 3--5 $\times 10^8$ K. At high energy, the component would turn over, consistent with early estimates of the peak temperature using {\it Ginga} \citep{brinkmann91,yuan95}. The plasma responsible for this X-ray emission, however, might originate from the disk corona rather than the jets. 

Furthermore, an estimate of the radiative cooling time scale it is substantially shorter than the flow time at the base of the jet. From Paper I, we expect a radiative cooling time, $t_r \equiv kT/(\Lambda[T]n_e$) $\sim 0.3$ s, whereas the flow time is $t_a \equiv r_j/v_j = 2.8$ s. We will revisit the validity of the adiabatic model in a forthcoming paper (Marshall et al. 2006). A model that may account for these dilemmas is one invoking substantially higher abundances. At the base of the jet where temperatures reach above 3$\times$10$^{7}$ K, cooling is dominated by bremsstrahlung processes. Consequently, if the Si abundance is increased to $\sim$10$\times$ solar, the continuum is depressed $\sim$10$\times$ and the cooling rate drops by $\sim$10$\times$. Thus, the cooling time would increase to $\approx$3 s. Coupled with a jet length decrease of 10$^{1/3}$, the flow time $t_a$ is below the radiative cooling time $t_r$ after abundance effect considerations. 

\subsection{Fluorescence}
\label{subsec:fl}

We observed two unresolved, unshifted line complexes likely due to fluorescence: one at 1.942$\pm$0.001 \AA\ (Fe~{\sc i}-{\sc xii}) and another at 7.131$\pm$0.006 \AA\ (Si~{\sc i}-{\sc v}). The line fluxes are substantial, with 1.0 $\times 10^{-4}$ ph cm$^{-2}$ s$^{-1}$ for Fe and 1.4 $ \times 10^{-5}$ ph cm$^{-2}$ s$^{-1}$ for Si, respectively. Adopting a distance of 4.85 kpc, the luminosities of these fluorescence lines are of the order \hbox{10$^{34}$ erg s$^{-1}$}. The stationary Fe~I line emission was previously observed by \cite{kotani96} at comparable strength. They excluded the possibility of a large companion, and consequently a substantial stellar wind, and they argued that the emission was likely due to partial illumination of an accretion disk by the central source. Given the moderate luminosity of the Fe~{\sc i} line, we suggest that this line emission originates near the companion, perhaps from clumps in a weakly ionized wind. Other high-mass X-ray binaries (HMXBs) like \hbox{Cyg X-1}, Vela X-1, Cen X-3, and GX 301-2 have been observed to produce line fluxes orders of magnitude higher, except during eclipse \citep{nagase,sako,patrick}. During occultation, the direct view of the compact object is obscured and the emission lines are attributed to dense material in the ionized winds rather than from the environment of the compact object. Likewise, the SS 433 fluorescence line luminosities are comparable to the quiescent fluorescent line luminosity in 4U 1700-37 \citep{boroson}. Consequently, the fluorescence line luminosities in SS 433 are similar to those of HMXBs during eclipse \citep{schulz,boroson}. The winds in these sources seem to have clumps of dense neutral and partially ionized material leading to fluorescence from Si~{\sc i} up to Si~{\sc viii} ions (Schulz et al. 2002, Boroson et al. 2003). Si~{\sc vii} and {\sc viii} ions have higher yields than Si {\sc i} ions. The absense of lines from these higher charge states indicates that the fluorescence in SS~433 originates from a much less ionized environment.

\subsection{Inferring the Binary Geometry and Black Hole Mass}
\label{subsec:bhmass}

We assume the system has the geometry shown in Fig.~\ref{fig:geometry} and approximate that the companion fully eclipses the receding jet for $r_j > x_j$. Our eclipse model requires that the disk is wind-fed and the companion star does not fill its Roche lobe radius $R_L$ which is considered unlikely by some theorists (for example, King, Taam, \& Begelman 2000). 

To examine this constraint of our model, we relate $R_L$ to the semimajor axis $a$ of the binary orbit and the mass ratio $q$ of the central object to the companion star as described by Ritter (1988):

\begin{equation}
\label{eq:roche}
R_L = \frac{0.49 q^{-2/3} a}{0.6 q^{-2/3} + \ln (1 + q^{-1/3})}      .
\end{equation}

\noindent
Utilizing the geometry shown in Figure~\ref{fig:distances}, the radius $R$ of the companion is

\begin{equation}
\label{eq:radius}
R = (a - d) \cos i   
\end{equation}

\noindent
where $d$ is the distance from the source of the jets to the point of intersection of the line of sight and the orbital plane and $i$ is the angle of inclination. If the companion fills its Roche lobe radius, then $R_L$ = $R$, and $d$ can be expressed in terms of $a$, $i$, and $q$:

\begin{equation}
\label{eq:newd}
d = a \bigg[1 - \frac{0.49 q^{-2/3} \sec i}{0.6 q^{-2/3} + \ln (1 + q^{-1/3})} \bigg] . 
\end{equation}

\noindent
With $i$=78.83\arcdeg\ from the kinematic model (Margon \& Anderson 1989), \hbox{$d$ $>$ 0} implies \hbox{$q$ $>$ 13.5}. Reasonable values of $q$ range 0 $<$ $q$ $<$ 1.5, so our model requires the companion star does not fill its Roche lobe. 

Using the angles intrinsic to the system in Fig.~\ref{fig:distances}, $d$ can be calculated:

\begin{equation}
\label{eq:distance}
d = \frac{x_j \sin \alpha}{\sin(90 - i)}                       .
\end{equation}

\noindent
With $x_{j}$ from $\S$\ref{sec:models}, $d=(3.5 \pm 0.8) \times 10^{11}$ cm. For a circular orbit, semimajor axis $a$ can be determined from the observable quantities $K_x$ and $K_o$, the semiamplitude velocities of the compact object and companion, respectively:

\begin{equation}
\label{eq:a}
a = \frac{P(K_x + K_o)}{2 \pi \sin i} = \frac{P K_x (1 + q)}{2 \pi \sin i}
\end{equation}
 
\noindent
where $q \equiv \frac{M_x}{M_o} = \frac{K_o}{K_x}$, and $M_x$ and $M_o$ are the masses of the compact object and companion, respectively.  With $K_x$= 175$\pm$20 km s$^{-1}$  \citep{fabrika} and $K_o$ = 100$\pm$15 km s$^{-1}$ \citep{gies1}, $a = (3.8\pm0.4) \times 10^{12}$ cm.  

Using Eq. 7, we find \hbox{$R = (6.7\pm0.7) \times 10^{11}$ cm}, or \hbox{$9.6\pm 1.0 R_{\sun}$}.  The uncertainty in the estimated radius is dominated by the uncertainty in the semimajor axis because $d$ and its uncertainty are rather small compared to $a$.  The value of $R$ is approximately one-third of the Roche lobe radius of the companion. Therefore, if our model is correct, the companion does not fill the Roche lobe, and accretion probably occurs from a stellar wind, as might be expected from an early type star. Given the small radius of the companion, we assume it is a main sequence star. Consequently, we can estimate its mass based on its radius. 

From an empirical mass-radius relation for main-sequence stars (Demircan \& Kahraman, 1991), a radius of \hbox{9.6$\pm1.0 R_{\sun}$} corresponds to a mass of \hbox{$35 \pm 7 ~ M_{\sun}$}. A star this massive can have a very strong stellar wind, consistent with our model that the companion is smaller than the Roche lobe. Using $K_x$ and $K_o$ from above, we find that the mass of the compact object is \hbox{$20 \pm 5 ~ M_{\sun}$}. This argument indicates the compact object responsible for the relativistic jets is a black hole. We note that the validity of this result depends on the adequacy of the adiabatic model and the assumption that the red and blue jets have similar velocities and are co-axial. These issues will be explored in our subsequent paper on SS 433 (Marshall et al. 2006). 

In recent work, Hillwig et al. 2004 obtained different values for the mass ratio $q$ (and consequently semimajor axis $a$) than the one used above. Based on weak absorption features in the optical spectra, they predict the companion is an evolved A-star with a mass $\approx$10.9 $\pm$3.1 $M_{\sun}$ and the compact object is a black hole with mass $\approx2.9 \pm 0.7 M_{\sun}$. This classification of the donor star is difficult because of the bright accretion disk that swamps signatures of the companion. Our model requires accretion to occur through a wind process, so the companion cannot be an A supergiant by our analysis. Future high-resolution optical and X-ray spectral observations of SS 433 are necessary to distinguish the true nature of the donor star. 

To check the robustness in our determination of the companion's radius, we calculate $R$ for different values of $q$. \cite{gies2} estimate $q = 0.72$ and \hbox{$a = 5.92\times10^{12}$ cm}. Using Eq.~\ref{eq:radius}, we find \hbox{$R = 1.08 \times10^{12}$ cm = 15.50 $R_{\sun}$} in this case. \cite{hillwig} obtain \hbox{$q = 0.26$} and \hbox{$a = 3.08\times10^{12}$ cm}, which corresponds to \hbox{$R = 5.29 \times10^{11}$ cm = 7.60 $R_{\sun}$}. $R$ $\approx$ 7.6--15.5 $R_{\sun}$ produces companion mass values between 25--80 $M_{\sun}$, and the resulting compact object mass is 6--60 $M_{\sun}$.

\subsection{Future Work} 

This work generates numerous questions we will address in upcoming papers. Foremost, our results rely on the assumptions that the adiabatic model adequately fits the observed \hbox{X-ray} spectra and that the jets have identical velocities and directions. As discussed in $\S$\ref{sec:models} and $\S$\ref{subsec:adiabatic}, the adiabatic model has shortcomings, and we plan to reexamine the jet cooling using recent 200 ks HETGS observation of SS 433 (ObsIDs 5512, 5513, 5514, 6360). Additionally, the observed line broadening may arise from the divergence of conical outflow or from Doppler shift variations through the course of an observation (see $\S$3). We plan to constrain the origins of this broadening in our next paper as well. Further optical measurements are needed in order to determine $K_o$ and $K_x$ more definitively. These velocities are vital to firmly establish the nature of SS 433. We are currently completing simultaneous observations of SS 433 in the X-ray, optical, and radio bands in order to examine the jets and their changes as they move outward from the disk in a more comprehensive manner. 

\acknowledgements

This work was partially supported by NASA contract NAS8-38249, Smithsonian Astrophysical Observatory contract SV3-73016 for the Chandra X-ray Center, and a National Science Foundation Graduate Research Fellowship (LAL). 


\clearpage

\begin{figure}
\epsscale{0.8}
\plotone{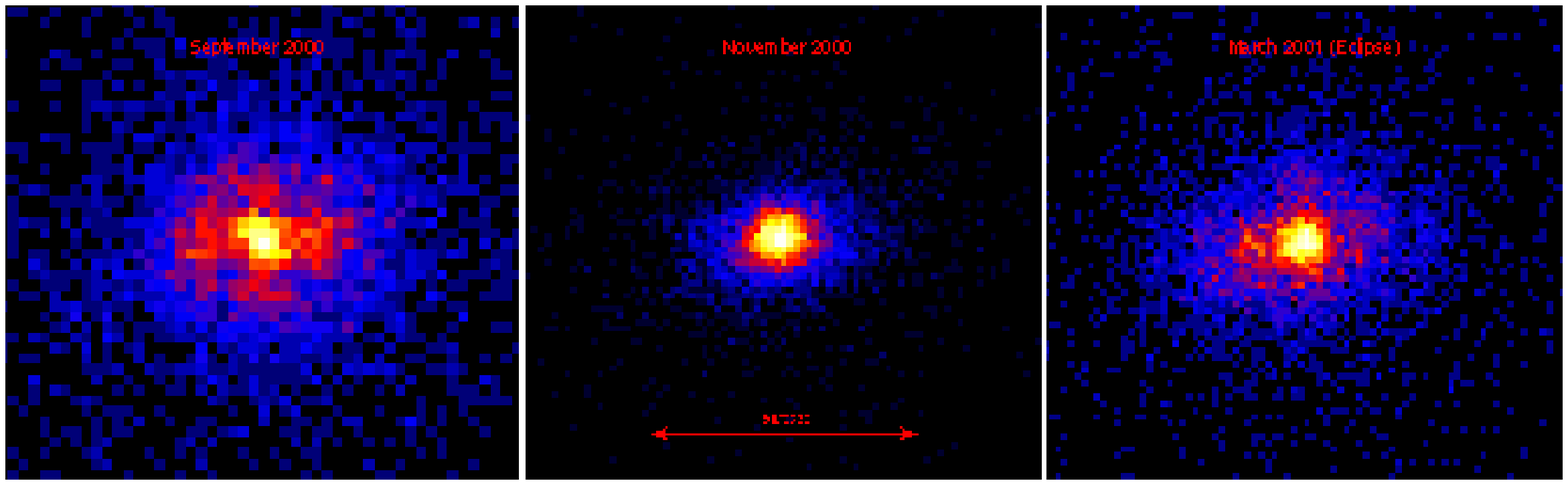}
\caption{Images of SS 433 from the zeroth order of the {\em Chandra} HETGS.   The images are extended along the east-west direction on a scale of 2--5\arcsec\ as discovered by \cite{marshall02} [Paper I]. The extent is comparable to that observed in the radio band by \cite{hjellming}. The left image was taken when the eastern jet neared maximal blueshift (Paper I). In the second observation (middle), the jet precessed around to point away from the observer, making it more redshifted than the western jet.  Finally, the March 2001 observation was taken at the same precessional phase as the first image, but the source was eclipsed.  Possible N-S structure in the first observation was not confirmed in later images.  The point source was weakest in November 2000.  Note that the point source is still very bright during eclipse, indicating that the innermost, hottest regions of the jets are not occulted.}
\label{fig:image}
\end{figure}  

\begin{figure}
\epsscale{0.8}
\plotone{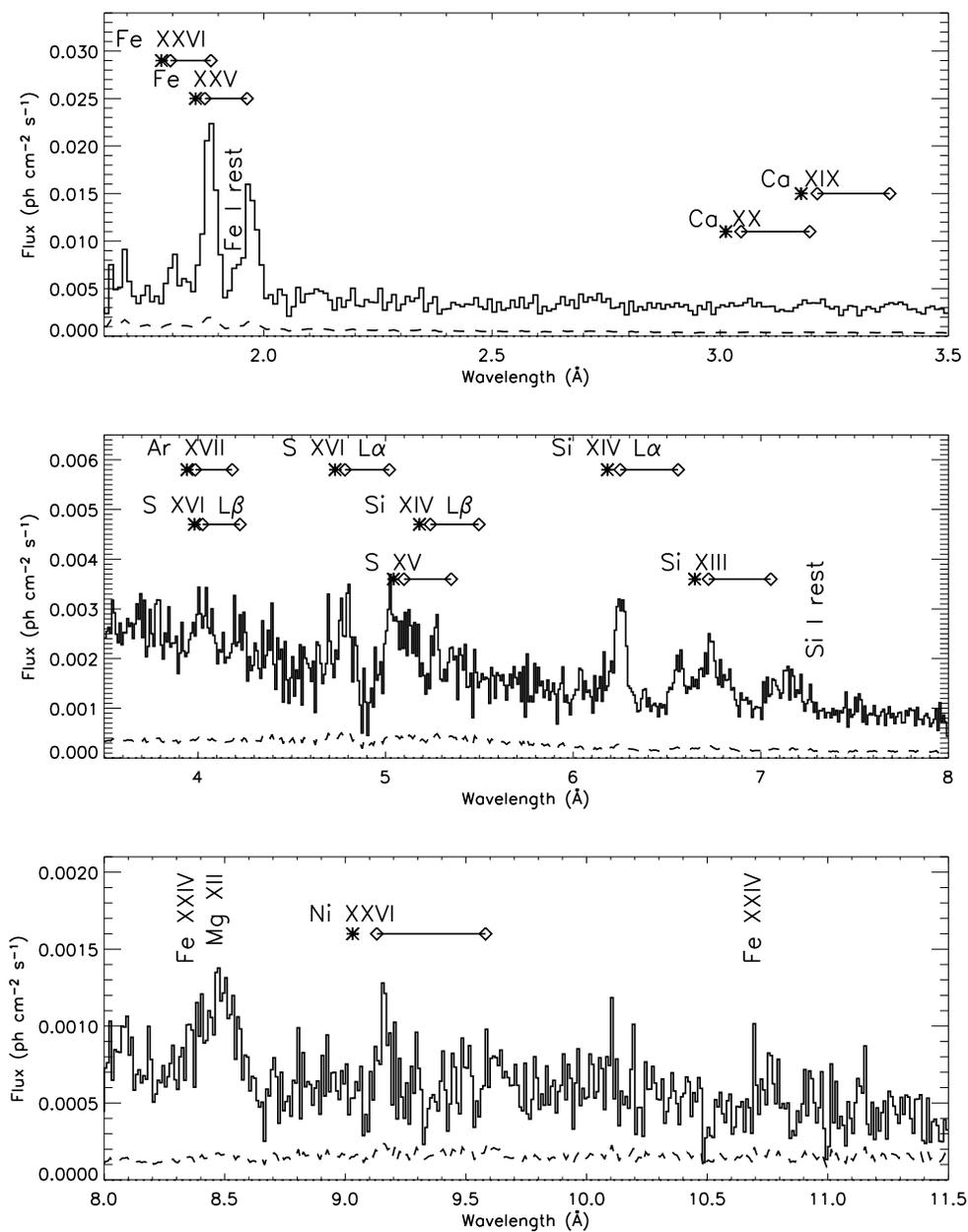}
\caption{The X-ray spectrum of SS 433 observed with the {\em Chandra} HETGS in March 2001. The HEG and MEG data were combined at the resolution of the MEG spectrum. Line identifications are shown and measurements are given in Table~\ref{tab:linefluxes}. Horizontal lines connect the locations of blue- and red-shifted lines (diamonds) to the rest wavelengths (asterisks). The dashed line gives the statistical uncertainties. Emission lines are resolved and their strengths indicate the plasma is collisionally dominated.  The Fe {\sc xxv} line from the approaching jet is confused with the Fe {\sc xxvi} line of the receding jet.}
\label{fig:spectrum}
\end{figure}

\begin{figure}
\epsscale{0.8}
\plotone{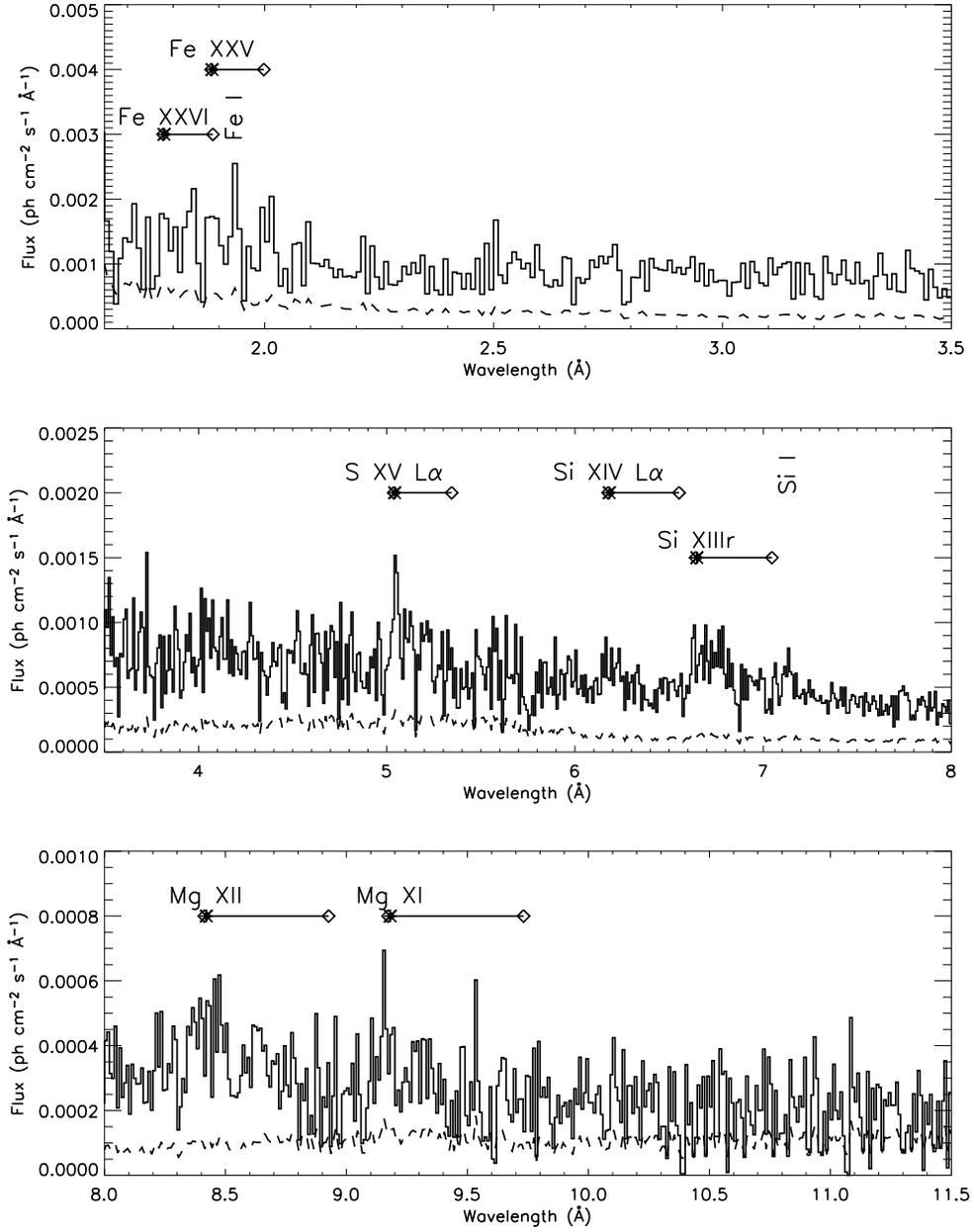}
\caption{The X-ray spectrum of SS 433 observed with the {\em Chandra} HETGS in November 2000, shown as in Fig.~\ref{fig:spectrum}.  Possible line identifications are shown.  Lines were extremely weak during this observation in both the red jet ($z_r = 0.070$) and blue jet ($z_b = -0.001$).  The overall flux level is very low compared to the observation reported in Paper I and the March 2001 observation (Fig.~\ref{fig:spectrum}).}
\label{fig:spectrum_obs2}
\end{figure}

\begin{figure}
\epsscale{0.8}
\plotone{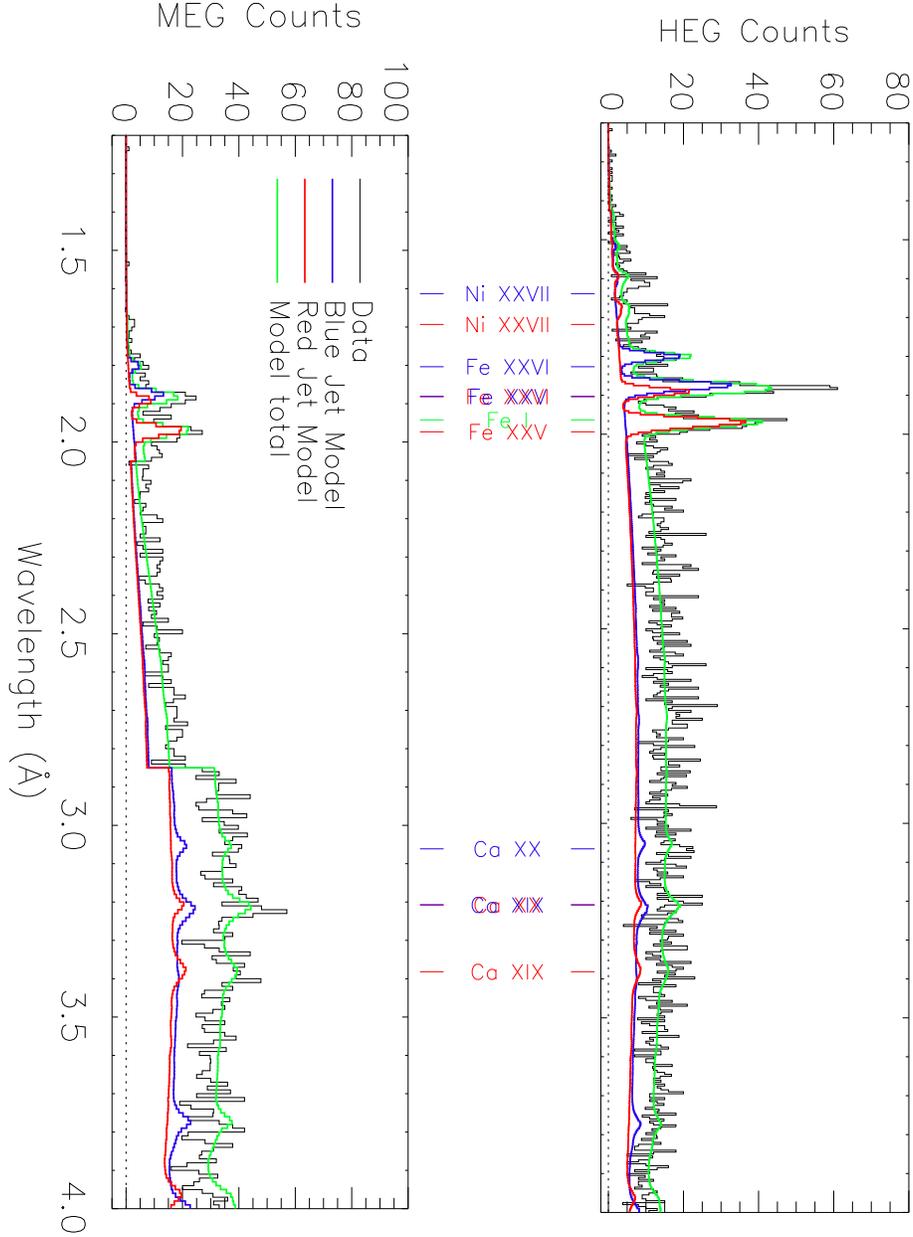}
\caption{The 1.2--4.0 \AA\ portions of the HEG and MEG spectra of SS 433 observed with the Chandra HETGS, compared to models of the spectra of the blue and red jets. The sum of the red and the blue spectra give the green curve. Line identifications are shown and measurements are given in Table~\ref{tab:linefluxes}. The continuum is dominated by thermal bremsstrahlung emission. The edge in the spectrum at 2.8 \AA\ is the result of excising data near a detector chip gap. In comparison to the first observation, the emission lines are more redshifted. The Fe {\sc i} line, at rest in the observed frame, is in the same location. The blue and red jet Fe {\sc xxv} and Fe {\sc xxvi} lines are of nearly equal strength in the model, indicating that the hottest portions of the jets are not blocked by the companion during eclipse.}
\label{fig:spectrum1} 
\end{figure}

\begin{figure}
\epsscale{0.8}
\plotone{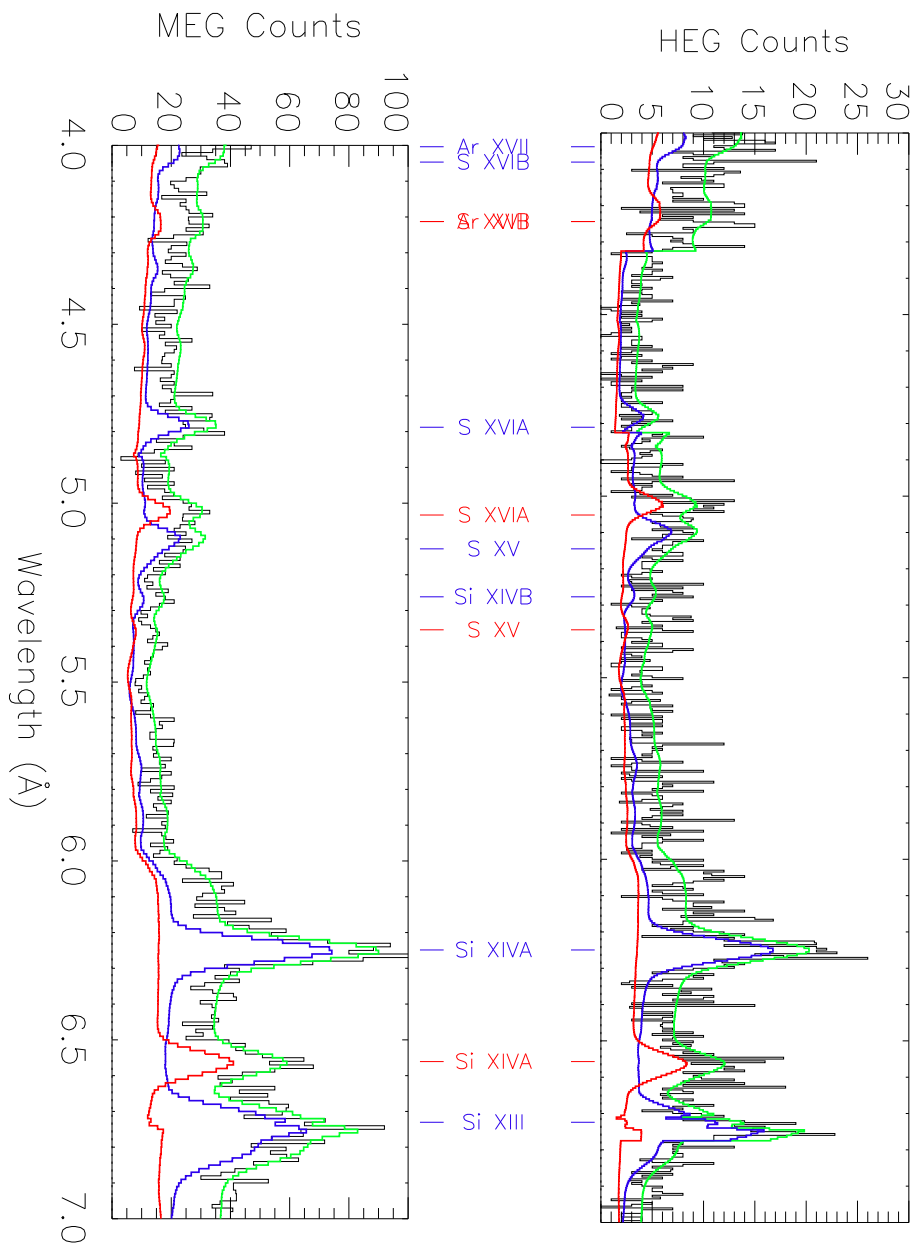}
\caption{Same as Figure~\ref{fig:spectrum1} except for the 4.0--7.0 \AA\ region. Unlike the first observation, the red jet is still strong in this portion of the spectrum. The overall model continuum is slightly higher than the data in the 4-6 \AA\ range, but the Si {\sc xiv} and Si {\sc xiii} lines are well fit.  Because the model includes the instrumental response, there is a steep rise near 6.1 \AA\ that results from the mirror Ir-M edge.}  
\label{fig:spectrum2} 
\end{figure}

\begin{figure}
\epsscale{0.8}
\plotone{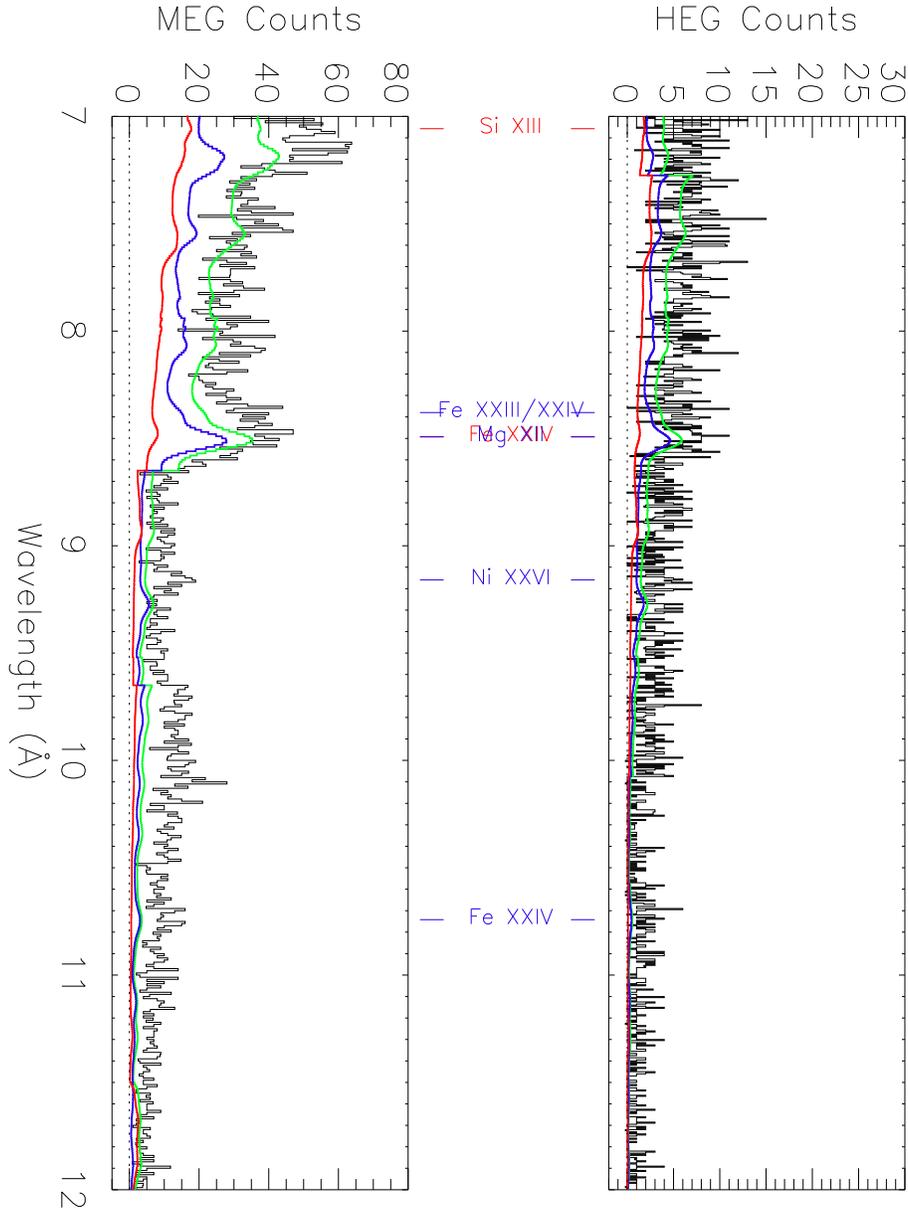}
\caption{Same as Figure~\ref{fig:spectrum1} except for the 7.0--12.0 \AA\ region. As in Paper I, the overall model continuum is lower than the data in this wavelength range. Additionally, fewer emission features can be distinguished. There is excess emission near the red Si {\sc xiii} line which can be attributed to a Si {\sc i} fluorescence line that is not included in the plasma model.}
\label{fig:spectrum3} 
\end{figure}

\begin{figure}
\epsscale{0.8}
\plotone{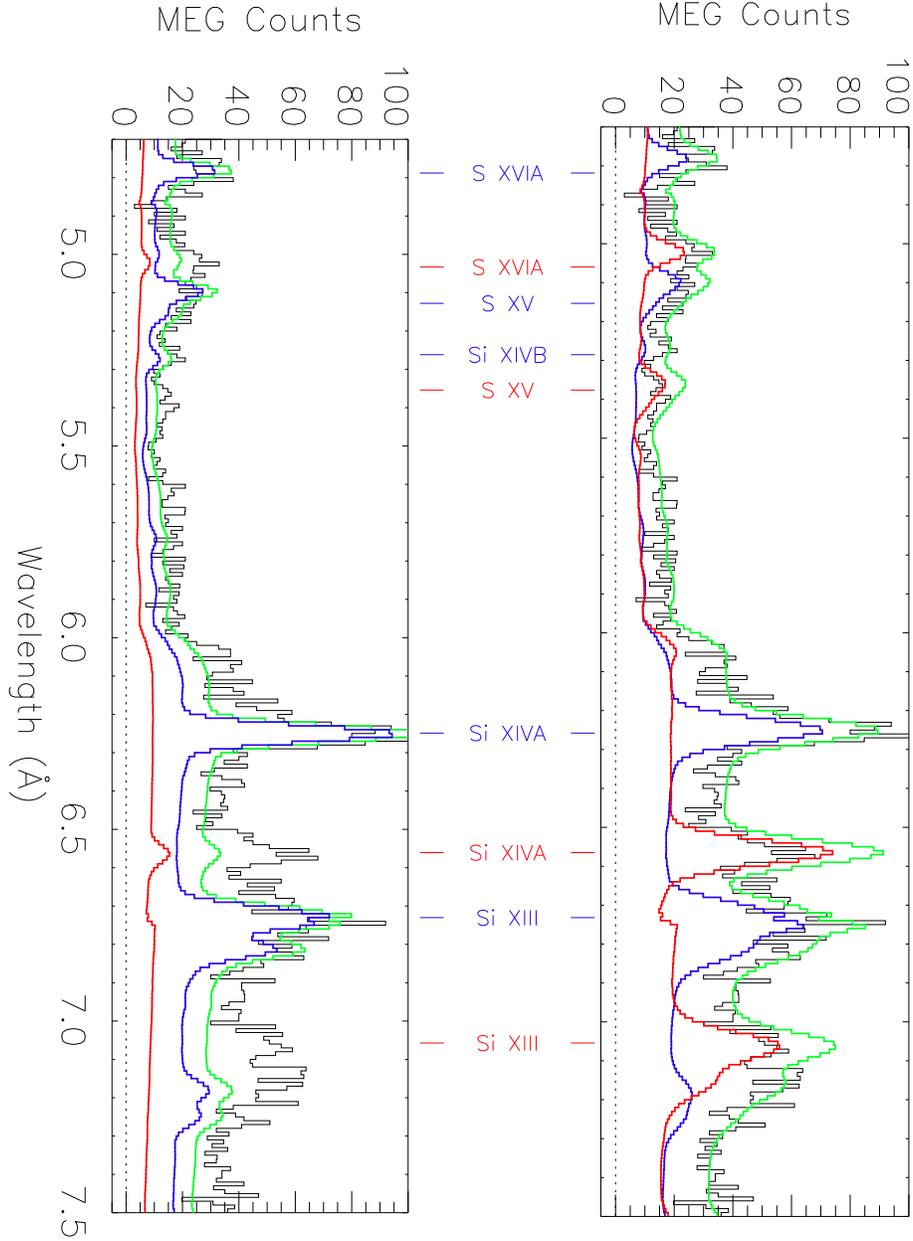}
\caption{Examples of poor fits obtained by varying the emission measures of zones two and three (see Table 3) in the red jet model. The models are color-coded as in Figure 4. {\em Top:} Spectra and models using the same emission measures for zones two and three in the red jet model as in the blue jet model. The S {\sc xv}, Si {\sc xiv}, and Si {\sc xiii} lines in the red jet are significantly overpredicted. {\em Bottom:} Same but with second and third emission measures set to zero in the red jet model. In this case, the S {\sc xv}, Si {\sc xiv}, Si {\sc xiii} lines are almost nonexistent in the red jet model.}
\label{fig:range}
\end{figure}

\begin{figure}
\epsscale{0.8}
\plotone{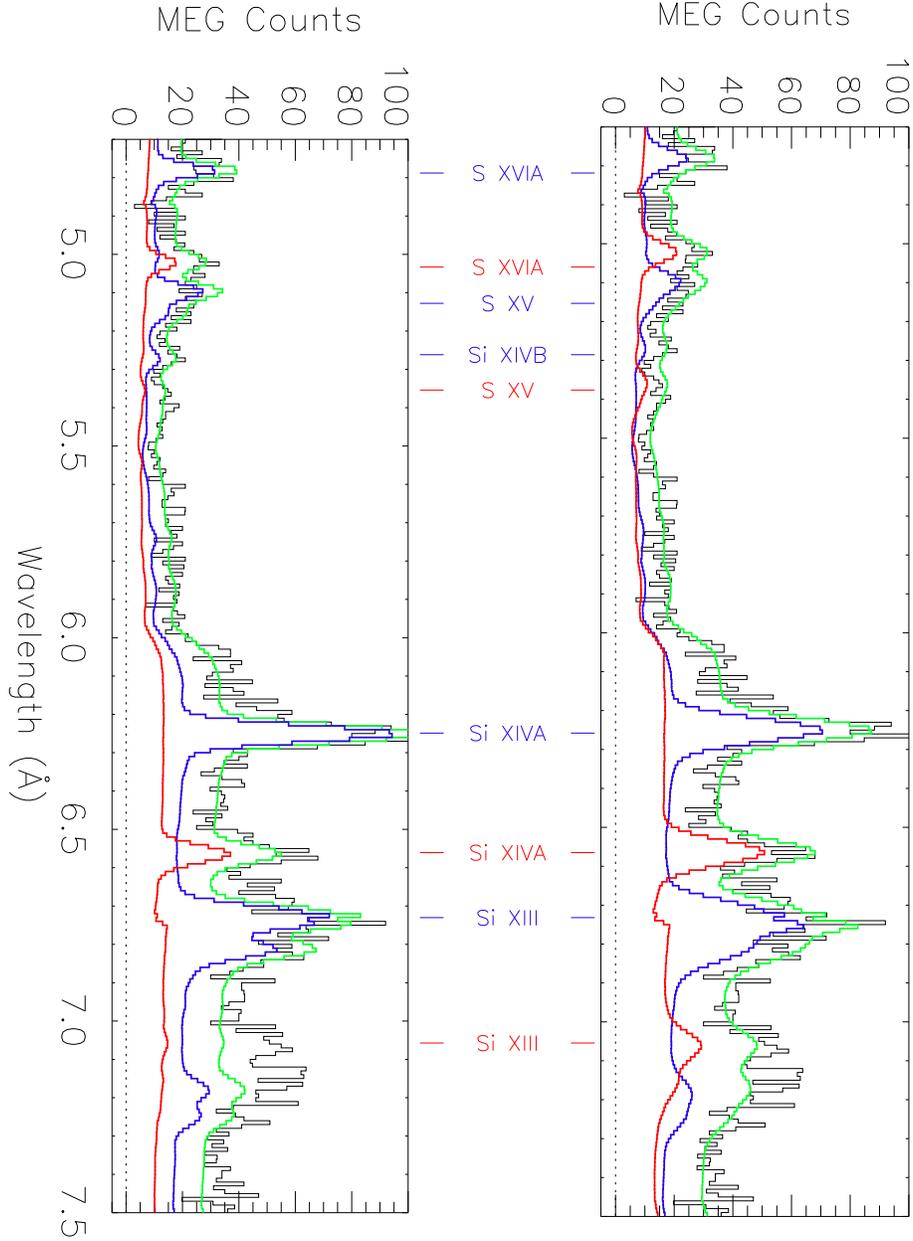}
\caption{Same as the previous figure except plots with reasonable fits to the data.{\em Top:} Plot with the red jet zones one and two set to zero and with zone three 70\% of the zone three blue jet value. Although slightly above the data, the Si {\sc xiv} and Si {\sc xiii} lines are reasonablely predicted by the model. This result gives a lower limit on the temperature at which the companion blocks the red jet. {\em Bottom:} Plot using 60\% of the second component normalization of the blue jet model while the third component normalization was set to zero. The model is a reasonable predictor for the observed lines (except around the Si fluorescence lines), and it places an upper limit on the temperature at which the companion blocks the red jet.}
\label{fig:confidence}
\end{figure}

\begin{figure}
\epsscale{0.8}
\plotone{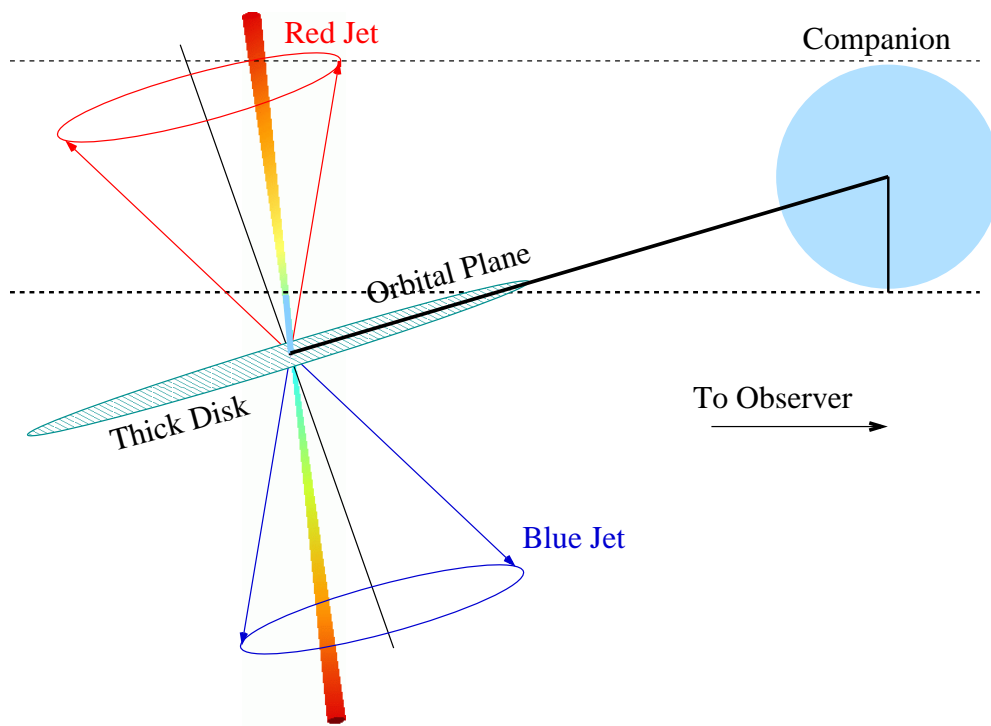}
\caption{Schematic geometry of the system. During eclipse, the companion blocks the cool, outer portion of the red jet from the observer but not the hotter part that is closer to the compact object.}
\label{fig:geometry}
\end{figure}

\begin{figure}
\epsscale{0.8}
\plotone{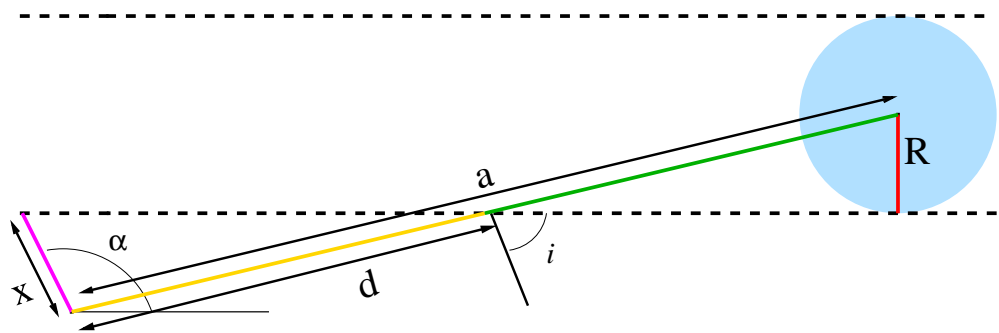}
\caption{Geometry, as in Fig.~\ref{fig:geometry}, with distances and angles labeled. Appealing to basic trigonometry and geometry, we can use the measured value of $x$ to find $R$, the radius of the companion. We calculate $\alpha$ in $\S$\ref{sec:linewidths}, the system inclination $i$ comes from the standard kinematic model \cite{ma89}, and the semimajor axis $a$ is derived from the system mass ratio and the size of the companion's Roche lobe (which may not be filled, see $\S$\ref{subsec:bhmass}). Thus, by measuring the distance $x_j$ using fits to the X-ray spectrum, we can determine the length $d$ and the stellar radius, $R$ ($\S$.~\ref{subsec:bhmass}).}
\label{fig:distances}
\end{figure}

\clearpage
\begin{deluxetable}{clccl}
\tablecolumns{5}
\tablewidth{0pc}
\tabletypesize{\scriptsize}
\tablecaption{SS 433 X-ray Emission Lines \label{tab:linefluxes} }
\tablehead{\colhead{$\lambda_{rest}$} & \colhead{$\lambda_{obs}$}
    & \colhead{$z$} & \colhead{Flux} & \colhead{Identification} \\
  \colhead{(\AA)} & \colhead{(\AA)} & \colhead{}
    & \colhead{($10^{-6}$ ph cm$^{-2}$ s$^{-1}$)} & \colhead{} }
\startdata
\cutinhead{Blue Jet}
1.780 & 1.804 $\pm$ 0.004 & 0.0133 $\pm$ 0.0020 & 116. $\pm$ 25. & Fe {\sc xxvi} Ly$\alpha^a$ \\
1.855 & 1.881 $\pm$ 0.009 & 0.0144 $\pm$ 0.0049 & 593. $\pm$ 37. & Fe {\sc xxv} 1s2p-1s$^2$ \tablenotemark{a}\\
3.020 & 3.061 $\pm$ 0.009 & 0.0136 $\pm$ 0.0029 & 33. $\pm$ 10. & Ca {\sc xx} Ly$\alpha$ \\
3.186 & 3.207 $\pm$ 0.007 & 0.0066 $\pm$ 0.0021 & 46. $\pm$ 10. & Ca {\sc xix} 1s2p-1s$^2$ \tablenotemark{b}\\
3.952 & 4.004 $\pm$ 0.007 & 0.0130 $\pm$ 0.0016 & 31. $\pm$ 10. & Ar {\sc xvii} 1s2p-1s$^2$ \\
3.991 & 4.047 $\pm$ 0.005 & 0.0140 $\pm$ 0.0014 & 105. $\pm$ 16. & S {\sc xvi} Ly$\beta$ \\ 
4.729 & 4.788 $\pm$ 0.005 & 0.0124 $\pm$ 0.0010 & 88. $\pm$ 14. & S {\sc xvi} Ly$\alpha$ \\
5.055 & 5.128 $\pm$ 0.006 & 0.0143 $\pm$ 0.0012 & 88. $\pm$ 13. & S {\sc xv} 1s2p-1s$^2$ \\
5.217 & 5.262 $\pm$ 0.007 & 0.0085 $\pm$ 0.0013 & 53. $\pm$ 12. & Si {\sc xiv} Ly$\beta$ \\
6.141 & 6.249 $\pm$ 0.002 & 0.0108 $\pm$ 0.0003 & 142. $\pm$ 8. & Si {\sc xiv} Ly$\alpha$ \\
6.675 & 6.730 $\pm$ 0.003 & 0.0083 $\pm$ 0.0004 & 88. $\pm$ 7. & Si {\sc xiii} 1s2p-1s$^2$ \\
8.310 & 8.379 $\pm$ 0.011 & 0.0083 $\pm$ 0.0014 & 22. $\pm$ 5. & Fe {\sc xxiii}/{\sc xxiv} \\
8.421 & 8.493 $\pm$ 0.005 & 0.0085 $\pm$ 0.0006 & 61. $\pm$ 6. & Mg {\sc xii} Ly$\alpha$ \tablenotemark{c}\\
9.103 & 9.166 $\pm$ 0.010 & 0.0071 $\pm$ 0.0011 & 25. $\pm$ 7. & Ni {\sc xxvi} \\
10.636 & 10.742 $\pm$ 0.013 & 0.0100 $\pm$ 0.0012 & 31. $\pm$ 6. & Fe {\sc xxiv} \\
\cutinhead{Rest frame}
1.937 & 1.942 $\pm$ 0.004 & 0.0025 $\pm$ 0.0018 & 100. $\pm$ 21. & Fe {\sc i} \\
7.128 & 7.131 $\pm$ 0.006 & 0.0004 $\pm$ 0.0008 & 14. $\pm$ 3. & Si {\sc i} - Si {\sc vii} \\
\cutinhead{Red Jet}
1.780 & 1.881 $\pm$ 0.001 & 0.0570 $\pm$ 0.0005 & 593. $\pm$ 37. & Fe {\sc xxvi} Ly$\alpha$ \tablenotemark{a}\\
1.855 & 1.973 $\pm$ 0.001 & 0.0641 $\pm$ 0.0006 & 348. $\pm$ 28. & Fe {\sc xxv} 1s2p-1s$^2$ \tablenotemark{a}\\
3.020 & 3.207 $\pm$ 0.007 & 0.0619 $\pm$ 0.0022 & 46. $\pm$ 10. & Ca {\sc xx} Ly$\alpha$ \tablenotemark{b}\\
3.186 & 3.382 $\pm$ 0.007 & 0.0614 $\pm$ 0.0021 & 41. $\pm$ 10. & Ca {\sc xix} 1s2p-1s$^2$ \\
3.952 & 4.213 $\pm$ 0.008 & 0.0659 $\pm$ 0.0020 & 57. $\pm$ 11. & Ar {\sc xvii} 1s2p-1s$^2$\\
3.991 & 4.213 $\pm$ 0.008 & 0.0557 $\pm$ 0.0020 & 57. $\pm$ 11. & S {\sc xvi} Ly$\beta$ \\
4.729 & 5.033 $\pm$ 0.004 & 0.0643 $\pm$ 0.0008 & 113. $\pm$ 13. & S {\sc xvi} Ly$\alpha$ \\
5.055 & 5.354 $\pm$ 0.013 & 0.0591 $\pm$ 0.0026 & 37. $\pm$ 12. & S {\sc xv} 1s2p-1s$^2$ \\
6.182 & 6.561 $\pm$ 0.004 & 0.0612 $\pm$ 0.0006 & 58. $\pm$ 6. & Si {\sc xiv} Ly$\alpha$ \\
6.675 & 7.057 $\pm$ 0.006 & 0.0573 $\pm$ 0.0010 & 32. $\pm$ 5. & Si {\sc xiii} 1s2p-1s$^2$ \\
7.989 & 8.493 $\pm$ 0.005 & 0.0630 $\pm$ 0.0007 & 61. $\pm$ 6. & Fe {\sc xxiv} \tablenotemark{c} \\
9.103 & 9.637 $\pm$ 0.013 & 0.0588 $\pm$ 0.0014 &  25. $\pm$  6. & Ni {\sc xxvi} \\
\enddata

\tablenotetext{a}{The red jet Fe {\sc xxvi} Ly $\alpha$ line is blended with the blue jet Fe {\sc xxv} line.}
\tablenotetext{b}{The red jet Ca {\sc xx} Ly $\alpha$ line is blended with the blue jet Ca {\sc xix} line.}
\tablenotetext{c}{The red jet Fe {\sc xxiv} line is blended with the blue Mg {\sc xii} Ly $\alpha$ line.}
\end{deluxetable}

\begin{deluxetable}{ll}
\tablecolumns{2}
\tablewidth{0pc}
\tablecaption{Line Widths
\label{tab:linewidths} }
\tablehead{\colhead{Wavelength Range} & \colhead{$v$\tablenotemark{a}} \\
   \colhead{(\AA)} & {(km s$^{-1}$)} }
\startdata
1.5 -- 2.5	&	1500 $\pm$  100 \\
2.5 -- 4.0	&	1600 $\pm$ 350 \\
4.0 -- 5.5	&	1600 $\pm$  ~200 \\
5.5 -- 7.5	&	1200 $\pm$  ~50 \\
7.5 -- 11.5 &	1200 $\pm$  ~100 \\
\enddata
\tablenotetext{a}{Velocity width ($\sigma$) of Gaussian line profiles.}
\end{deluxetable}

\begin{deluxetable}{cccccc}
\tablecolumns{6}
\tablewidth{0pc}
\tablecaption{Jet Parameters from a Multi-Temperature Model\tablenotemark{a}
\label{tab:model} }
\tablehead{ & & Blue Jet & Red Jet & & \\
\colhead{Zone} & \colhead{$T$} & \colhead{$EM$} & \colhead{$EM$} &\colhead{$r$} & \colhead{$n_e$} \\
 & (10$^6$ K) & (10$^{57}$ cm$^3$) & (10$^{57}$ cm$^3$) & (10$^{10}$ cm) & (10$^{14}$ cm$^{-3}$) }
\startdata
1 &   6.3 &  4.72  & 0 & 20.5  &  0.40 \\
2 &  12.6 &  4.60  & 0 & 12.2  &  1.00 \\
3 &  31.6 &  8.63  & 8.63  & 6.13 &  4.00 \\
4 & 126.  &  8.83  & 8.83  &2.17 & 20.0  \\
\enddata
\tablenotetext{a}{The interstellar absorption
column density was fit simultaneously with the $EM$ values
at each temperature.  The best fit absorption column was fixed at
$2.07 \times 10^{22}$ cm$^{-2}$.  The values of $r$ and $n_e$ are derived using the adiabatically expanding jet model determined in Paper I.}
\end{deluxetable}

\end{document}